\documentclass[letter]{aa} 
\usepackage{threeparttable}
\usepackage{graphicx}
\usepackage{amsmath}
\usepackage{comment}
\usepackage{txfonts}
\newcommand{\ued}[1]{{\textcolor{black}{#1}}}

\begin{document} 



\title{Jupiter's ``cold'' formation in the protosolar disk shadow} 

\subtitle{An explanation for the planet's uniformly enriched atmosphere}

\author{Kazumasa Ohno
          \inst{1,2}
          \and
          Takahiro Ueda
          \inst{3}
          }

   \institute{Department of Astronomy and Astrophysics, University of California, 1156 High St, Santa Cruz, CA 95064, US\\
  \email{kono2@ucsc.edu}
   \and
   Department of Earth and Planetary Sciences, Tokyo Institute of Technology, Meguro, Tokyo, 152-8551, Japan
        \and
    National Astronomical Observatory of Japan, Osawa, Mitaka, Tokyo, 181-8588, Japan\\
             \email{takahiro.ueda@nao.ac.jp}
             }

\date{}

\abstract
   {Atmospheric compositions offer valuable clues to planetary formation and evolution. Jupiter has been the most well-studied giant planet in terms of its atmosphere; however, the origin of the Jovian atmospheric composition remains a puzzle as the abundances of nitrogen and noble gases as high as those of other elements could only originate from extremely cold environments.}
   {We propose a novel idea for explaining the Jovian atmospheric composition: Dust pileup at the H$_2$O snow line casts a shadow and cools the Jupiter orbit so that N$_2$ and noble gases can freeze. Planetesimals or a core formed in the shadowed region can enrich nitrogen and noble gases as much as other elements through their dissolution in the envelope.
   }
   {We compute the temperature structure of a shadowed protosolar disk with radiative transfer calculations. Then, we investigate the radial volatile distributions and predict the atmospheric composition of Jupiter with condensation calculations.}
   {We find that the vicinity of the current Jupiter orbit, approximately $3$--$7~{\rm AU}$, could be as cold as $\lesssim30~{\rm K}$ if the small-dust surface density varies by a factor of $\gtrsim30$ across the H$_2$O snow line. 
   According to previous grain growth simulations, this condition could be achieved by weak disk turbulence if silicate grains are more fragile than icy grains.
   The shadow can cause the condensation of most volatile substances, namely N$_2$ and Ar.
   We demonstrate that the dissolution of shadowed solids can explain the elemental abundance patterns of the Jovian atmosphere even if proto-Jupiter was formed near Jupiter's current orbit.}
  {The disk shadow may play a vital role in controlling atmospheric compositions. The effect of the shadow also impacts the interpretation of upcoming observations of exoplanetary atmospheres by JWST.}


\maketitle

\section{Introduction}

Atmospheric compositions encapsulate the information on planet formation.
Jupiter has been the most well-studied gas giants in terms of atmospheric composition.
Extensive observations by the Galileo probe, Cassini, and JUNO {spacecraft} have revealed that O, C, N, S, P, Ar, Kr, and Xe are all uniformly enriched from protosolar abundances by a factor of two to four \citep[][]{Mahaffy+00,Wong+04,Fletcher+09,Bolton+17,Li+20}.
The heavy-element enrichment likely originates from planetesimals and/or pebbles dissolved in the atmosphere \citep[e.g.,][]{Pollack+86,Iaroslavitz&Podolak07,Hori&Ikoma11,Venturini+16} and/or core erosion \citep[e.g.,][]{Guillot+04,Moll+17}.
Intriguingly, the abundances of highly volatile elements, such as N and Ar, are comparable to the other elemental abundances.

The origin of the uniformly enriched Jovian atmosphere has been a long-standing puzzle because solid dissolution hardly enhances the highly volatile elements.
\citet{Owen+99} suggested that the uniform enrichment originates from planetesimals formed in extremely cold ($<30~{\rm K}$) environments, where noble gases can freeze.
\citet{Oberg&Wordsworth19} suggested that the uniform enrichment could be explained if the Jovian core had formed beyond the Ar snow line, presumably placed at $>30~{\rm AU}$ \citep[see also][]{Bosman+19}.
However, the migration of a core from $>30~{\rm AU}$ to $5~{\rm AU}$ is a rare occurrence, according to theoretical studies on core formation and migration \citep{Bitsch+15,Bitsch+19}.
Even if such a migration occurs, the core arrives at the current orbit at $>1~{\rm Myr}$ \citep{Alibert+05b,Bitsch+19}. 
This is in tension with the isotope dichotomy of meteorites, indicating that Jupiter may demarcate the inner and outer Solar System within $\sim1~{\rm Myr}$ \citep{Kruijer+17,Kruijer+20}.

In this letter we propose an alternative idea to explain the uniformly enriched Jovian atmosphere: The shadow cast by dust pileup at the H$_2$O snow line produced extremely cold environments near the current Jupiter orbit, leading to the formation of volatile-rich solids there. 
The dissolution of such ``shadowed solids'' can enrich the abundances of highly volatile elements as much as other elements, even if proto-Jupiter formed near the current orbit.

\section{Method}\label{sec:Method}

\subsection{Basic idea}
The mid-plane temperature in the outer region of protoplanetary disks is determined by the stellar irradiation grazing at the disk surface.
Since the stellar light comes from the center of the disk, a shadowed region where the direct stellar light never reaches is potentially generated, depending on the surface structure of the inner region.
For example, a puffed inner disk rim can block off stellar light and cast a shadow, so-called self-shadowing \citep{Dullemond+01,Dullemond&Dominik04}.
The presence of the self-shadow has been suggested from several disk observations \citep[e.g.,][]{Garufi+17,Avenhaus+18}.
The shadow can also emerge when dust grains are accumulated somewhere and block off the stellar light.
\citet{Ueda+19} showed that a dust pileup at the inner edge of the dead zone  casts a shadow behind it, producing cold regions of ${\sim}50~{\rm K}$ at ${\sim}2$--$7~{\rm AU}$ around a Herbig-type star.

One potential mechanism of the shadowing onto the current Jupiter orbit is a dust pileup at the H$_2$O snow line.
The dust surface density inside the H$_2$O snow line can be enhanced by orders of magnitude because efficient fragmentation slows the radial drift of silicate grains \citep{Birnstiel+10, Banzatti+15,Pinilla+17}.
With high surface density and scale height, such fragmented dust may cast a shadow behind the H$_2$O snow line and provide cold regions where volatile substances can freeze.
The primary purpose of this letter is to investigate how cold this shadowed region is and what kinds of volatile gases can freeze there.

\subsection{Numerical model}\label{sec:disk_model}
We adopted a parameterized disk model in which the dust surface density steeply varies around the H$_2$O snow line, described by
\begin{equation}\label{eq:sigmag}
\Sigma_{\rm g} = 670 \left( \frac{r}{1 \rm AU} \right)^{-3/5}~{\rm g~cm^{-2}},
\end{equation}
\begin{equation}\label{eq:sigmad}
\Sigma_{\rm d} = \left\{ \begin{array}{ll}
    0.01\Sigma_{\rm g} & (r< R_{\rm SL}) \\
    0.01f_{\rm SL}\Sigma_{\rm g} & (r\geq R_{\rm SL}),
  \end{array} \right.
\end{equation}
where $\Sigma_{\rm g}$ is the gas surface density, $\Sigma_{\rm d}$ is the surface density of the small dust (i.e., $\lesssim100~{\rm \mu m}$) contributing to the opacity, and $R_{\rm SL}$ is the radial location of the H$_2$O snow line set to $R_{\rm SL}=1.3~{\rm AU}$.
We varied the parameter $f_{\rm SL}$ to examine the effects of the dust pileup magnitude.
The actual value of $f_{\rm SL}$ depends on how {easily the fragmentation of silicate and icy grains takes place}, which is controlled by the turbulence strength and stickiness of the grains \citep{Birnstiel+10,Banzatti+15, Pinilla+17}.
For example, \citet{Banzatti+15} reported the surface density variations of $f_{\rm SL}\sim0.1$, $0.001$, and $0.01$ around the H$_2$O snow line for the turbulence strengths of $\alpha_{\rm t}={10}^{-2}$, ${10}^{-3}$, and ${10}^{-4}$, respectively.

With the radial density profile, we calculated the temperature structure of the shadowed region using the Monte Carlo radiative transfer code RADMC-3D \citep{RADMC}.
Since the vertical dust distribution and the temperature structure are mutually dependent, the radiative transfer calculations were iteratively performed to obtain the self-consistent disk structure \citep{Ueda+19}. 
The simulations also included the internal energy released by the disk accretion with a form of $q_{\rm acc}=(9/4)\alpha_{\rm t}\rho_{\rm g}c_{\rm s}^{2}\Omega_{\rm K}$ with $\alpha_{\rm t}=3\times10^{-4}$, where $\rho_{\rm g}$ is the gas density, $c_{\rm s}$ is the sound speed, and $\Omega_{\rm K}$ is the Keplerian frequency.
We adopted the DSHARP dust opacity model with the minimum and maximum dust radii of $0.1$ and $100~{\rm \mu m}$ (\citealt{Birnstiel+18}; see also \citealt{Henning96, Draine03, Warren08}).

{  
Using the computed temperature structure, we estimated the mid-plane volatile abundances in gas and solid phases at each orbital distance by balancing condensation and sublimation rates (Appendix \ref{Appendix:cond_model}).
Following \citet{Oberg&Wordsworth19}, we performed the calculation for H$_2$O, CO$_2$, CO, C$_2$H$_6$, N$_2$, NH$_3$, Ar, Kr, and Xe based on the up-to-date protosolar abundances in \citet{Asplund+21}.}
We then estimated the elemental abundance of the Jovian atmosphere, $\mathcal{N}_{\rm i}/\mathcal{N}_{\rm H}$, using volatile abundances of gas and solids at $5~{\rm AU}$, as (see Appendix \ref{Appendix:atmosphere})
\begin{equation}\label{eq:atm_calc}
    \frac{\mathcal{N}_{\rm i}}{\mathcal{N}_{\rm H}}=\frac{N_{\rm t,i}}{N_{\rm H}}\frac{N_{\rm g,i}}{N_{\rm t,i}}+ \frac{M_{\rm Z}}{M_{\rm env}}\frac{\overline{m}_{\rm H+He}}{\overline{m}_{\rm d}}\frac{\mathcal{N}_{\rm H}+\mathcal{N}_{\rm He}}{\mathcal{N}_{\rm H}}\frac{n_{\rm d,i}}{\sum{n_{\rm d,i}}},
\end{equation}
where $M_{\rm Z}/M_{\rm env}$ is the mass fraction of volatiles dissolved in the envelope, $\overline{m}_{\rm H+He}=1.24~{\rm amu}$ and $\mathcal{N}_{\rm H}/(\mathcal{N}_{\rm H}+\mathcal{N}_{\rm He})=0.92$ are the mean atomic mass and hydrogen fraction of solar composition gases, {$N_{\rm t,i}$ and $N_{\rm g,i}$ are the total and gas-phase number density of molecules for species i, $n_{\rm d,i}$ is the surface number density of the molecules adsorbed onto dust,} and $\overline{m}_{\rm d}=\sum{m_{\rm i}n_{\rm d,i}}/\sum{n_{\rm d,i}}$ is the mean molecular mass of the dissolved volatiles.
{  We used sulfur to evaluate $M_{\rm Z}/M_{\rm env}$ since most sulfur likely exists as refractory solids \citep{Kama+19}.}
We adjusted $M_{\rm Z}/M_{\rm env}$ so that the S abundance {  matches the observed S abundance of Jupiter} and then calculated other elemental abundances.
We also estimated the atmospheric composition of Saturn from solid- and gas-phase volatile abundances at $10~{\rm AU}$ {using the same procedure}.

\section{Results}\label{sec:Results}

\begin{figure}[t]
\centering
\includegraphics[clip, width=\hsize]{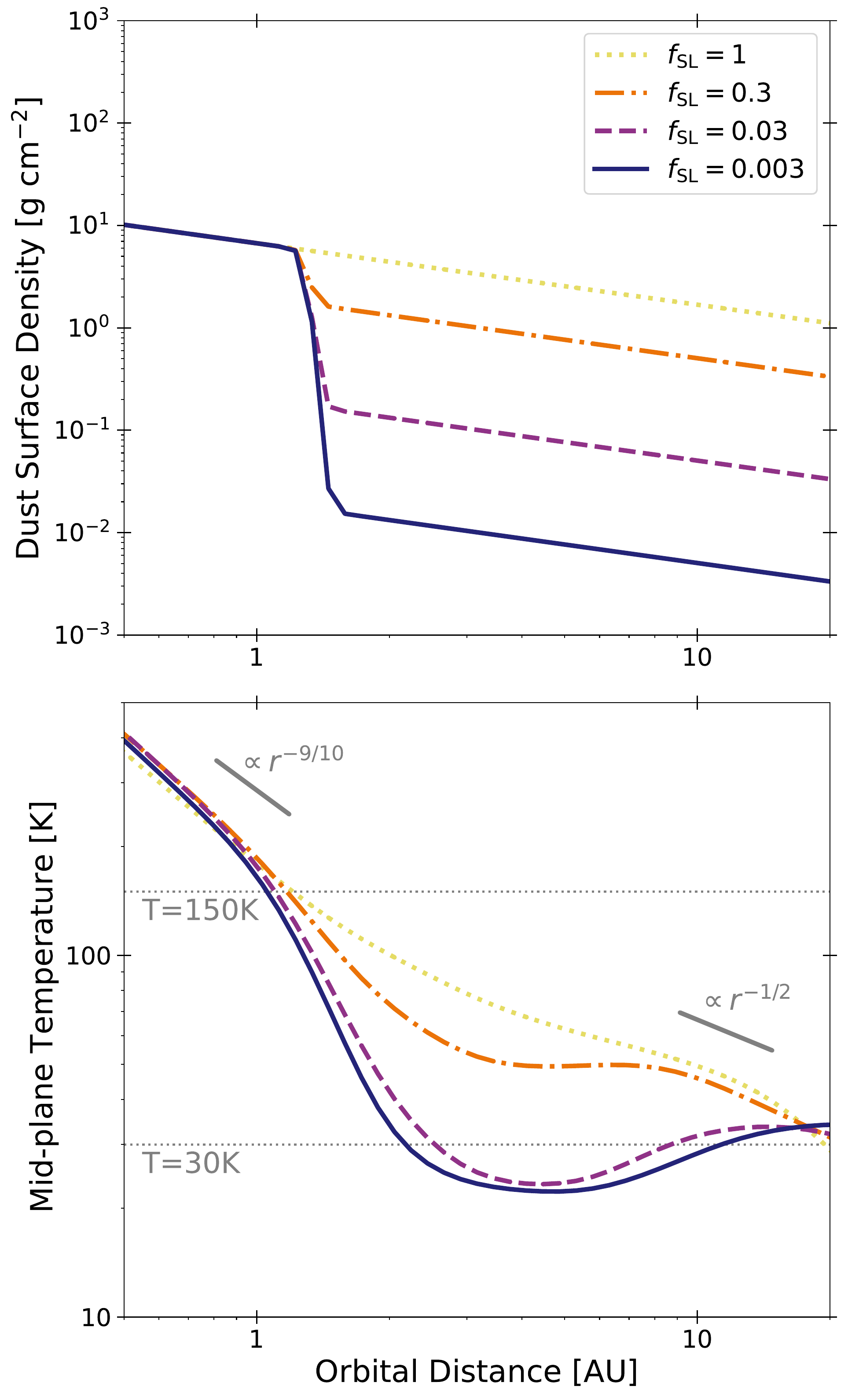}
\caption{
{Physical structures of a protoplanetary disk with the shadow cast by dust pileup at the H$_2$O snow line.} (Top) Assumed surface density profiles of small dust. (Bottom) Mid-plane temperature {at each orbital distance}. Different colored lines show the profiles for different $f_{\rm SL}$. }
\label{fig:shadow}
\end{figure}
\begin{figure*}[t]
\centering
\includegraphics[clip, width=\hsize]{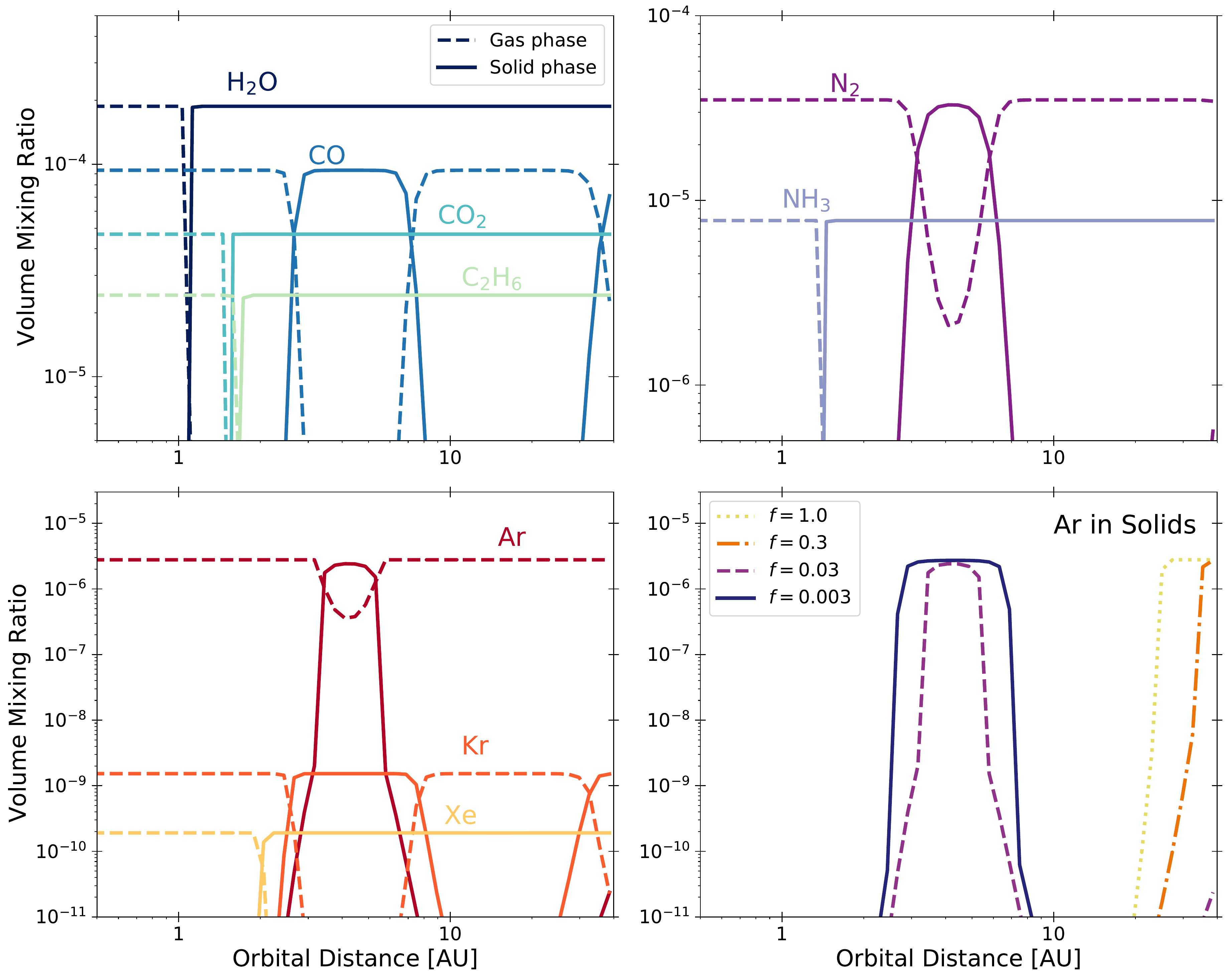}
\caption{Volume mixing ratio of volatile substances in solid and gas phases at each orbital distance. The bottom right panel shows the Ar abundance in solids for different $f_{\rm SL}$. We set $f_{\rm SL}=0.03$ in the remaining panels. 
}
\label{fig:cond}
\end{figure*}
\begin{figure*}[t]
\centering
\includegraphics[clip, width=\hsize]{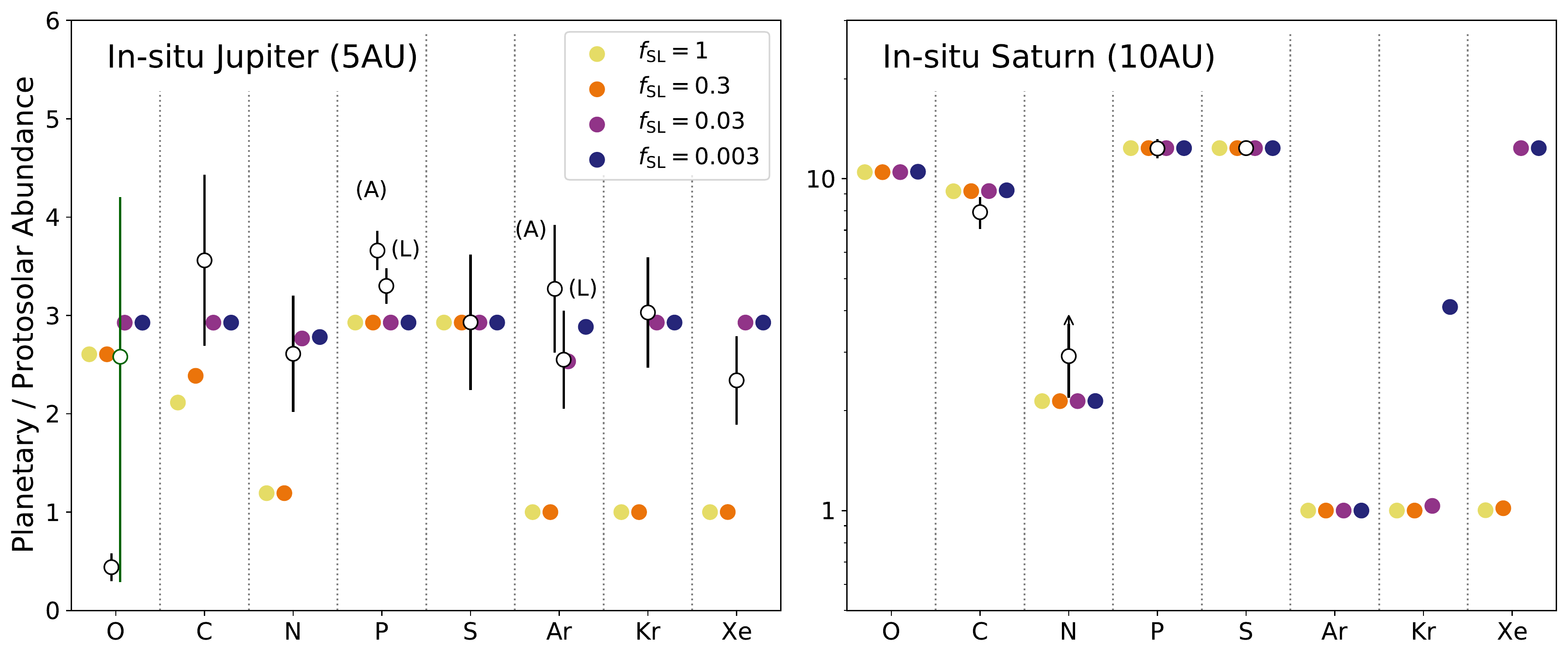}
\caption{Predicted elemental abundances of the Jovian (left) and Saturnian  ({right}) atmospheres normalized by the protosolar value. 
We have evaluated atmospheric elemental abundances using the volatile abundances at $5~{\rm AU}$ for Jupiter and  $10~{\rm AU}$ for Saturn.
The empty dots denote the abundances constrained by the Galileo probe, Cassini, and JUNO observations from \citet[][]{Atreya+20}.
The observational abundances are normalized by the protosolar abundances of \citet{Asplund+21}, and we also plot the abundances normalized by the protosolar abundances of \citet{Lodders+09} for P and Ar, {  where (A) and (L) indicate \citet{Asplund+21} and \citet{Lodders+09}}, respectively.
The empty green dot denotes the oxygen abundance at the deep equatorial region of Jupiter retrieved by JUNO \citep{Li+20}. 
We caution that the observed P abundances may not represent the bulk abundances, that the N abundance in Saturn may be a lower limit, and that the S abundance in Saturn is model dependent \citep[see][]{Atreya+16}.
}
\label{fig:summary}
\end{figure*}

\subsection{Mid-plane temperature in a shadowed disk}
Figure \ref{fig:shadow} shows the mid-plane temperature for different $f_{\rm SL}$.
In general, the temperature inside the H$_2$O snow line is independent of $f_{\rm SL}$ and is primarily controlled by viscous heating, yielding the radial dependence of $\propto r^{-9/10}$ \citep[e.g.,][]{Oka+11}.
The temperature structures with different $f_{\rm SL}$ converge to a similar flared disk solution of $T\propto r^{-1/2}$ at $\ga10~{\rm AU}$ \citep{Kenyon&Hartmann87}, implying that these regions are directly irradiated by sunlight.

The dust pileup at the H$_2$O snow line drastically alters the mid-plane temperature at ${\sim}2$--$10~{\rm AU}$ owing to shadow formation.
The temperature at ${\sim}2$--$10~{\rm AU}$ decreases with decreasing $f_{\rm SL}$, as the shadow extends to more outer regions.
Even a small change in dust surface density with $f_{\rm SL}=0.3$ causes the cooling of $\sim20~{\rm K}$ from a non-shadowed disk with $f_{\rm SL}=1$.
Remarkably, the shadow cools the vicinity of the current Jupiter orbit, $\sim3$--$7~{\rm AU}$, to $<30~{\rm K}$ for $f_{\rm SL}<0.03$.
Although the dust-to-gas mass ratio inside the H$_2$O snow line may be higher than that assumed here, especially in the young disks of $\la1~{\rm Myr}$ \citep[e.g.,][]{Birnstiel+10}, we have confirmed that the shadow can still cool the Jupiter orbit to $<30~{\rm K}$ in that case (Appendix \ref{Appendix:young_disk}).

\subsection{Radial volatile distributions}
Most volatile substances freeze in the cold shadowed regions.
Figure \ref{fig:cond} shows the radial distributions of volatile abundances in solids and gases for $f_{\rm SL}=0.03$.
In the shadowed disk, CO$_2$, C$_2$H$_6$, and NH$_3$ have snow lines at nearly the same orbit, ${\sim}1.5~{\rm AU}$. 
These crowded snow lines originate from the steep temperature gradient in the shadow right behind the H$_2$O snow line.
CO, N$_2$, Ar, and Kr exhibit intriguing orbital distributions: They freeze into solids at $\sim3$--$7~{\rm AU}$ but return to gas phases at $\ga7~{\rm AU}$.
These volatile distributions are significantly different from conventional non-shadowed disks \citep[e.g.,][]{Oberg+11,Oberg&Wordsworth19}.

The magnitude of the dust pileup at the H$_2$O snow line determines if volatile substances can freeze in shadowed regions.
The bottom right panel of Fig. \ref{fig:cond} shows the solid-phase abundance of Ar, the most volatile substance in our model.
For $f_{\rm SL}=1$ and $0.3$, Ar only freezes at outer regions of $>20~{\rm AU}$.
On the other hand, Ar freezes in solids at $\sim3.5$--$6~{\rm AU}$ for $f_{\rm SL}=0.03$ and at $\sim2.5$--$8~{\rm AU}$ for $f_{\rm SL}=0.003$.
Thus, highly volatile substances can freeze even at the current Jupiter orbit if the small-dust surface density varies by a factor of $\ga30$ across the H$_2$O snow line.

\subsection{Comparison with Jovian atmospheric composition}
Figure \ref{fig:summary} compares the estimated elemental abundances of the Jovian atmosphere with those constrained by observations. 
In weakly shadowed disks of $f_{\rm SL}\ge0.3$, N and noble gases have almost protosolar abundances as they are still present in gas phases at $\sim5~{\rm AU}$, {inconsistent with the observations}.
On the other hand, N and noble gases can be enriched as much as other elements for $f_{\rm SL}\le0.03$ as the cold shadowed regions cause the condensation of $\rm N_{\rm 2}$ and noble gases at $5~{\rm AU}$.
The estimated dissolved volatile mass is $M_{\rm Z}/M_{\rm env}=0.022$ for $f_{\rm SL}\le0.03$, equivalent to $\approx 7M_{\rm \oplus}$ if we approximate $M_{\rm env}{\approx}M_{\rm p}$.
This is well below the total heavy element mass of ${\sim}8$--$36M_{\rm \oplus}$, the upper limit allowed for dissolved volatile mass, constrained by JUNO \citep{Wahl+17,Ni19}.
The estimated abundance pattern  explains the uniform enrichment observed in the Jovian atmosphere well.

In our shadow scenario, Saturn,  in contrast to Jupiter, may not undergo the uniform enrichment.
This is because Saturn's orbit is near the outer edge of the shadowed regions; this area is too warm to allow the condensation of $\rm N_{\rm 2}$ and noble gases, although this depends on the gas and dust disk properties.
As a result, N and Ar abundances become significantly lower than the others.
The estimated elemental abundances explain the C, N, P, and S abundances of Saturn well, although the N, P, and S abundances are still uncertain \citep[see][]{Atreya+16}.
A possible depletion of N and noble gases is intriguing because other uniform enrichment scenarios, such as the accretion of volatile-enriched disk gases \citep[e.g.,][]{Guillot&Hueso06,Monga&Desch15,Ali-Dib17,Mousis+19}, would likely result in the uniform enrichment on Saturn being similar to that on Jupiter.
Since noble gas abundances can only be constrained by in situ measurements,  future entry probe missions on Saturn \citep[][]{Mousis+16,Simon+18} would help to distinguish the shadow scenario from other scenarios.

\section{Discussion}\label{sec:Discussion}
\subsection{Implications for Jupiter formation}
There are two possible pathways to incorporate shadowed solids into the Jovian atmosphere: core dissolution and solid accretion after envelope formation.
In the former case, the shadow scenario suggests that the Jovian core may form near the current orbit via planetesimal \citep[e.g.,][]{Pollack+96,Inaba+03,Kobayashi&Tanaka18} and/or pebble accretion \citep[e.g.,][]{Lambrechts&Johansen12,Lambrechts+14}.
The upward mixing of the dissolved core could enrich the envelope if primordial composition gradients exist \citep[][]{Guillot+04,Vazan+18a}.
The JUNO observations do suggest the presence of a dissolved core \citep{Wahl+17,Ni19}.
In the latter case, the core could be formed elsewhere.
Based on isotopes in meteorites, \citet{Alibert+18} suggested that substantial planetesimal accretion occurred on Jupiter during the envelope accretion.
Conversely, the capture of planetesimals is inefficient for gap-opening Jupiter, and an enhanced surface density of planetesimals may have been needed to achieve sufficient enrichment \citep{Shibata&Ikoma19}.
It is vital to understand how many planetesimals form 
in the shadowed regions.

\subsection{Implications from disk observations}
Recent observations of protoplanetary disks have shown that shadow-like structures are common in scattered light images \citep{Avenhaus+18,Garufi+18}.
The origin of these structures is still unclear, but most of them are observed at the outer region ($>$10 AU) and hence would not be related to the H$_2$O snow line.
Due to the limited angular resolution and the bright emission coming directly from the central star, it is still challenging to constrain the disk surface structure within 10 AU.
However, the prevalence of the shadow on the disk surface implies that the disk temperature no longer follows the simple power-law profile.

Interestingly, recent multiwavelength studies {of the Atacama Large Millimeter/submillimeter Array (ALMA)} have shown that ring and gap structures reside even in the optically thick region (\citealt{Carrasco+19,Macias+21}).
This implies that the substructures are induced by the temperature variation and/or intensity reduction from scattering (e.g., \citealt{baobab19,Zhu+19,Sierra+20,Ueda+20}), not by the density variation.
For example, the HL Tau disk has a gap at $\sim$13 AU at the wavelength where the region is expected to be optically thick \citep{Carrasco+19}. 
This gap might be linked with a shadow cast by the H$_2$O snow line, which is expected to be located at $\sim$ 10 AU.
However, the dust size in the inner region of the HL Tau disk is unlikely to be smaller than that at $>$10 AU (\citealt{Carrasco+19,Ueda+21}).
To precisely estimate the dust size and the disk temperature, multiwavelength millimeter observations with more than four wavelengths are needed, which have only been conducted for a few disks.
The future multiwavelength ALMA observations would tell us if the gap is associated with the shadow.

\subsection{Diagnostic ratios for exoplanet observations}\label{appendix:C}
\begin{figure}[t]
\centering
\includegraphics[clip, width=\hsize]{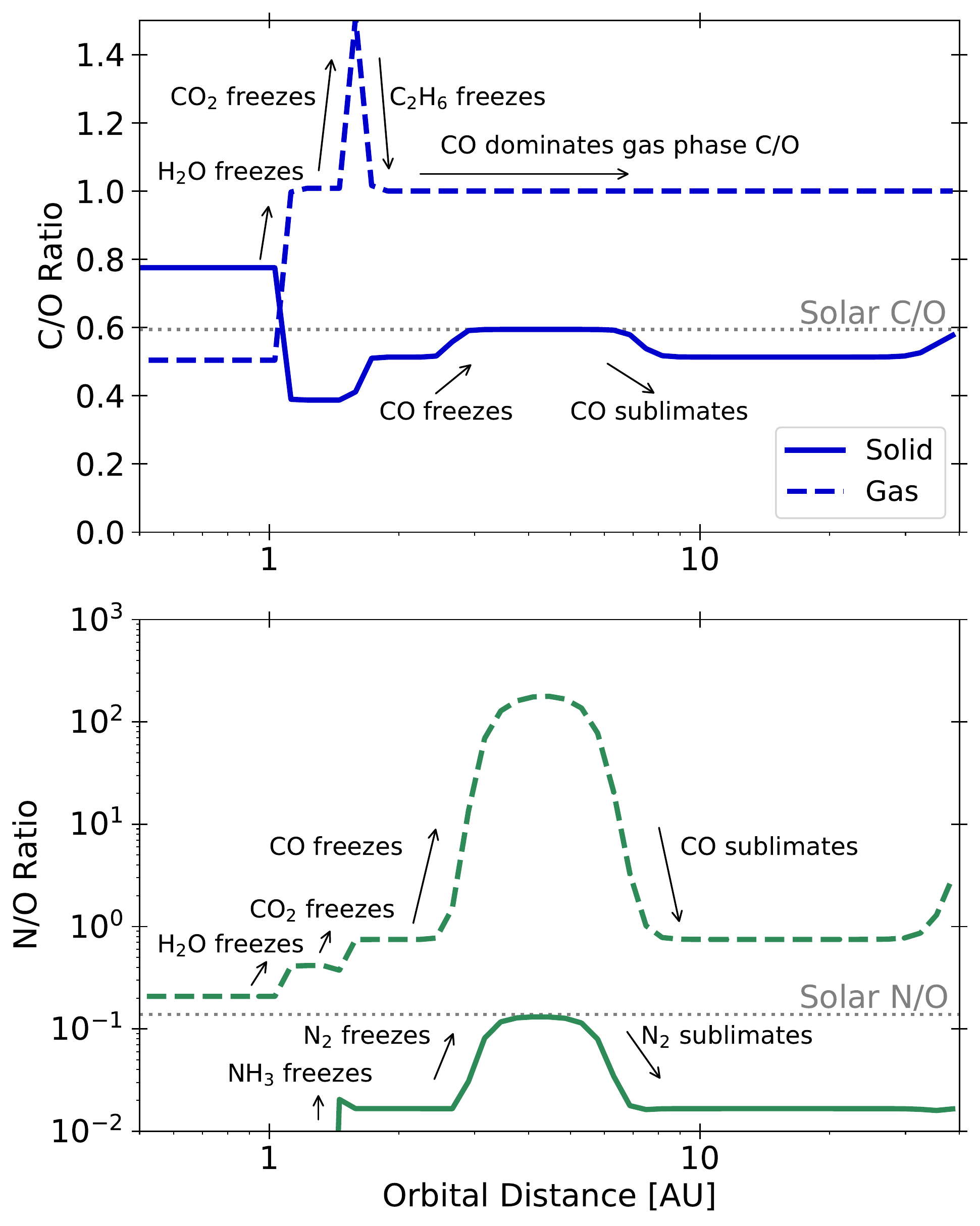}
\caption{C/O and N/O ratios as a function of orbital distances in the shadowed disk. The solid and dashed lines show the ratios in solid and gas phases, respectively. We set $f_{\rm SL}=0.03$.}
\label{fig:C/O}
\end{figure}

It is generally expected 
that the C/O ratios of exoplanetary atmospheres are linked to the compositions of protoplanetary disks \citep[e.g.,][]{Oberg+11,Madhusudhan+14,Mordasini+16,Booth+17,Cridland+20,Notsu+20,Miley+21}.
In shadowed disks, the solid C/O ratio reaches the solar value in shadowed regions, {similar to the} outer disk (Fig. \ref{fig:C/O}).
The gas C/O ratio is almost unity at $\ga2~{\rm AU}$ because CO accommodates most of the gas-phase C and O there.
{Thus, planets formed in the shadowed regions would have atmospheres with C/O ratios similar to those formed in the outer parts of protoplanetary disks.}

It is worth noting that the N/O ratio shows a spatial variation that is much larger than that of the C/O ratio (Fig. \ref{fig:C/O}).
The solid N/O ratio reaches a solar value in the shadowed region as all N and O reservoirs freeze.
Interestingly, the gas N/O ratio exceeds unity in the shadowed region because N$_2$ has a desorption energy lower than that of CO, leading to a higher N$_2$ vapor abundance.
Thus, the planets formed in the shadow would have atmospheres with solar N/O and super-solar metallicity (the case of Jupiter) or N/O$>1$ and sub-solar metallicity.
Exoplanetary nitrogen abundances can be constrained by the observations of NH$_3$ and HCN \citep{Macdonald&Madhusudhan17a,Macdonald&Madhusudhan17}.
These molecules would be detectable by the upcoming observations of {the James Webb Space Telescope (JWST)}. 

\section{Summary}\label{sec:Summary}

The disk shadow may significantly affect disk temperature structures and planetary compositions.
If the dust surface density varies by a factor of $\ga30$ (i.e., $f_{\rm SL}\la0.03$) across the H$_2$O snow line, the vicinity of the current Jupiter orbit could be shadowed and as cold as $\la30~{\rm K}$. 
We have demonstrated that the uniformly enriched Jovian atmosphere could be explained if the Jovian core was formed in the shadowed regions or if the envelope experienced shadowed solid dissolution.
The required condition of $f_{\rm SL}\la0.03$ was observed for $\alpha_{\rm t}\la{10}^{-3}$ in previous grain growth simulations that assumed {the} fragmentation velocity to be $1$ and $10~{\rm m~s^{-1}}$ for silicate and H$_2$O ice grains \citep{Birnstiel+10,Banzatti+15,Pinilla+17}. 
Such a weak turbulence may be compatible with recent ALMA observations of several protoplanetary disks \citep[e.g.,][]{Pinte+16,VanBoekel+17,Flaherty+20}.
The past cold environment can be compatible with ammoniated rocks discovered on several Jupiter Trojan asteroids, Jupiter-family comets, and (1) Ceres \citep{deSanctis+15,Brown16,Poch+20}.
Although these indications of cold environments can be explained by the dynamical scattering of objects from the outer Solar System \citep[e.g.,][]{Morbidelli+05}, the shadow scenario might provide an alternative explanation.

\begin{figure}[t]
\centering
\includegraphics[clip, width=\hsize]{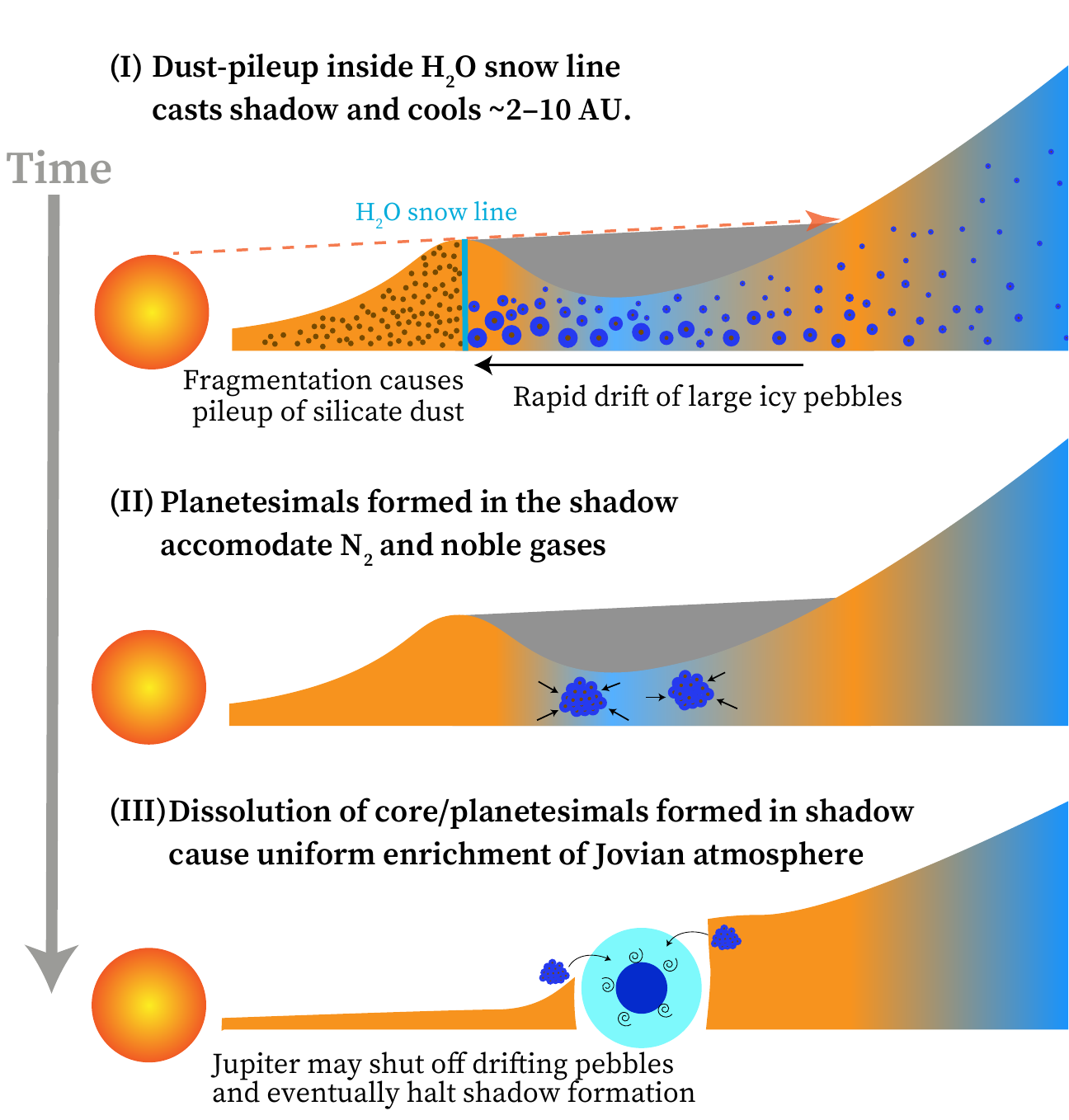}
\caption{Cartoon summarizing our shadow scenario for a uniform enrichment of the Jovian atmosphere.}
\label{fig:cartoon}
\end{figure}

Figure \ref{fig:cartoon} schematically summarizes our shadow scenario.
In reality, since the location of the H$_2$O snow line evolves over time \citep[e.g.,][]{Garaud&Lin07,Oka+11,Drazkowska&Dullemond18,Mori+21}, the shadow location would evolve as well.
The shadow may also disappear once Jupiter opens up a gap and shuts off drifting pebbles \citep{Morbidelli+16}.
Further modeling on time-evolving disks with the shadow is warranted to understand how the shadow impacts planetary formation and compositions, which will be the subjects of our future studies.

\section*{Acknowledgements}
We are grateful to {the} referee, Alexander Cridland, for insightful comments that greatly improved the quality of this paper.
We also thank Jonathan Fortney for helpful suggestions on the paper draft, Mario Flock for insightful comments on the numerical models, and Satoshi Okuzumi, Shota Notsu, and Hideko Nomura for fruitful conversations.
K.O. is supported by JSPS KAKENHI Grant Numbers JP19K03926 and JSPS Oversears Research Fellowship.
T.U. is supported by JSPS KAKENHI Grant Numbers JP19J01929.
Numerical computations were in part carried out on Cray XC50 at Center for Computational Astrophysics, National Astronomical Observatory of Japan.

\bibliographystyle{aa}

\appendix
{ 
\section{Elemental abundances}\label{Appendix:abundance}
\begin{table}[t]
 \centering
\begin{threeparttable}
  \caption{Elemental abundances in the protosun, Jupiter, and Saturn.}\label{table:0}
  \begin{tabular}{l r r r r} \hline
   Element & Protosun\tnote{a} & Jupiter/Protosun\tnote{b} &  Saturn/Protosun\tnote{b} \\ \hline \hline
    O/H & $5.62\times{10}^{-4}$ & $0.44\pm0.14$ & N/A \\
    &  & $2.6^{+2.3}_{-1.6}$ \tnote{c} & N/A \\
    C/H & $3.34\times{10}^{-4}$ & $3.56\pm0.87$ & $7.93\pm0.87$\\
    N/H & $7.78\times{10}^{-5}$ & $2.61\pm0.59$ & $>2.92\pm0.73$\\
    S/H & $1.52\times{10}^{-5}$ & $2.93\pm0.69$ & $12.37$\\
    P/H & $2.95\times{10}^{-7}$ & $3.66\pm0.20$ & $12.34\pm0.81$ \\
    Ar/H & $2.78\times{10}^{-6}$ & $3.27\pm0.65$ & N/A\\
    Kr/H & $1.53\times{10}^{-9}$ & $3.03\pm0.56$ &  N/A \\
    Xe/H & $1.91\times{10}^{-10}$ &  $2.34\pm0.45$ & N/A \\
    \hline
    \end{tabular}
        \begin{tablenotes}
        \raggedright
        \item[a] From Table B.1 of \citet{Asplund+21}.
        \item[b] From Table 1 of \citet{Atreya+20}. See also references therein.
        \item[c] From \citet{Li+20}. 
    \end{tablenotes}
    
   \end{threeparttable} 
\end{table}
Recently, \citet{Asplund+21} provided new recommended compositions of the protosun, which we have adopted in this study.
Table \ref{table:0} summarizes the up-to-date elemental abundances of the protosun along with the abundances of Jupiter and Saturn normalized by them.
We note that the protosolar abundances are higher than the photospheric abundances of the current Sun \citep[see Sect. 5 of][]{Asplund+21}.

\section{Condensation model}\label{Appendix:cond_model}
\subsection{Model description}
Here we describe the condensation model we used for evaluating volatile distributions.
The abundances can be evaluated by equating condensation and sublimation rates \citep[e.g.,][]{Oberg&Wordsworth19}.
The former is given by
\begin{equation}\label{eq:R_cond}
    R_{\rm cond,i}=\frac{\sigma_{\rm gr}}{4}N_{\rm g,i}v_{\rm th,i},
\end{equation}
where $\sigma_{\rm gr}$ is the total grain surface area per unit volume calculated from the dust size distribution used in the radiative transfer model, $N_{\rm g,i}$ is the number density of gas-phase atoms or molecules, and $v_{\rm th,i}=\sqrt{8k_{\rm B}T/\pi m_{\rm i}}$ and $m_{\rm i}$ are the mean thermal velocity and mass of species i atoms or molecules, respectively.
We note that the gas-phase number density is associated with the total number density, $N_{\rm t,i}$, as
\begin{equation}
    N_{\rm t,i}=N_{\rm g,i}+\sigma_{\rm gr}n_{\rm d,i},
\end{equation}
where $n_{\rm d,i}$ is molecular or atomic surface number density on grain surfaces.

The sublimation behavior depends on how much the molecules or atoms are adsorbed onto the grain surface.
In a low surface density regime where all adsorbed particles are exposed, the sublimation rate is proportional to the surface density, so-called first-order desorption \citep[e.g.,][]{Collings+03}.
As the surface density increases, two factors affect the sublimation rate.
The first is that only particles in a few of the uppermost molecular or atomic layers can directly desorb from the grain surface, leading to the upper limit of ``desorption active'' surface density to be set at $n_{\rm d,i}\sim{10}^{15}~{\rm cm}^{-2}$ \citep[e.g.,][]{Collings+03}.
The second is that the desorption energy decreases with increasing surface coverage of molecules or atoms owing to the increased number of particles weakly bound on the surface \citep{Fayolle+16,He+16}.
In the limit of the high surface density of adsorbed molecules or atoms, one can expect that the sublimation rate eventually approaches the rate of pure ice, which has been studied in the context of cloud microphysics in planetary atmospheres \citep[e.g.,][]{pruppacher&Klett97}. 

Based on this consideration, we modeled the sublimation rate as
\begin{equation}\label{eq:R_sub}
   R_{\rm subl,i}=\sigma_{\rm gr}\min{\left[\nu_{\rm i}n_{\rm d,i} \exp{\left(-\frac{E_{\rm des,i}}{k_{\rm B}T}\right)},\frac{P_{\rm s,i}}{4k_{\rm B}T}v_{\rm th,i} \right]},
\end{equation}
where $E_{\rm des}$ and $\nu$ are the desorption energy and attempt frequency of particles adsorbed onto grains and $P_{\rm s}$ is the vapor pressure.
Equation \eqref{eq:R_sub} returns to the first-order desorption rate at low $n_{\rm d,i}$ \citep[e.g.,][]{Collings+03} and ensures that the sublimation rate approaches that of pure ice in the limit of high $n_{\rm d,i}$ \citep[e.g.,][]{pruppacher&Klett97}.
The vapor pressure of each substance is taken from \citet{Fray&Schmitt09_vapor-pressure}.

Our condensation model assumes that the dust and gas share the same temperature, which is mostly valid at the dense disk mid-plane.
In the disk mid-plane, which the high-energy stellar photons hardly penetrate, gas particles tend to have lower temperatures than dust.
The gas heating rate by collisions with dust is given by \citep[e.g.,][]{Tielens05}
\begin{equation}
    m_{\rm g}n_{\rm gas}c_{\rm p}\frac{dT_{\rm gas}}{dt}=2n_{\rm gas}n_{\rm gr}\pi a^2 v_{\rm g}k_{\rm B}(T_{\rm gr}-T_{\rm gas})\alpha_{\rm a}, 
\end{equation}
where $m_{\rm g}$ is the mean mass of gas particles, $c_{\rm p}$ is the specific heat capacity, $\pi a^2$ is the mean cross section of dust, $v_{\rm g}=\sqrt{8k_{\rm B}T_{\rm gas}/\pi m_{\rm g}}$ is the mean thermal velocity, $\alpha_{\rm a}\sim0.15$ is the accommodation coefficient, $n_{\rm gas}$ and $n_{\rm gr}$ are the number densities of gas and dust, and $T_{\rm gr}$ and $T_{\rm gas}$ are the dust and gas temperatures.
The timescale with which the gas temperature is relaxed to the dust temperature, $T_{\rm dust}(dT_{\rm gas}/dt)^{-1}$, is given by
\begin{eqnarray}
    \nonumber
    \tau_{\rm relax}&\approx&\frac{1}{2n_{\rm gr}\pi a^2v_{\rm g}(1-\gamma^{-1})\alpha_{\rm a}}=\frac{\pi\rho_{\rm d}a\Omega^{-1}}{3f_{\rm d}\Sigma_{\rm g} (1-\gamma^{-1})\alpha_{\rm a}}\\
    &\sim& \frac{{10}^{-4}}{f_{\rm d}}~{\rm yr}~\left( \frac{a}{1~{\rm \mu m}}\right)\left( \frac{\Sigma_{\rm g}}{100~{\rm g~{cm}^{-2}}}\right)^{-1} \left( \frac{r}{5~{\rm AU}}\right)^{3/2}.
\end{eqnarray}
We used $(4\pi a^3\rho_{\rm d}/3)n_{\rm gr}=f_{\rm d}\Sigma_{\rm g}/(\sqrt{2\pi}c_{\rm s}\Omega^{-1})$ and $k_{\rm B}/m_{\rm g}c_{\rm p}=(1-\gamma^{-1})$, where $\gamma=7/5$ is the adiabatic index for diatomic gases, $\rho_{\rm d}$ is the dust internal density, $\Omega$ is the Kepler frequency, and $f_{\rm d}$ is the dust-to-gas mass ratio. 
We assumed $\rho_{\rm d}=3~{\rm g~{cm}^{-3}}$ and the solar mass.
The lowest dust-to-gas mass ratio is $f_{\rm d}=3\times{10}^{-5}$ in this study.
Even in this case, the relaxing timescale is $\sim 3~{\rm yr}$, still much shorter than the timescale of disk evolution.
}

\begin{table}[t]
 \centering
\begin{threeparttable}
  \caption{Summary of material properties.}\label{table:1}
  \begin{tabular}{l r r r r} \hline
   Species & $N_{\rm t}/N_{\rm H}$ \tnote{a} & $E_{\rm des}/k_{\rm B}~[{\rm K}]$\tnote{b} & $\nu_{\rm }~[{\rm s^{-1}}]$ \\ \hline \hline
    H$_2$O & $1.87\times{10}^{-4}$ & $5600$ & ${10}^{15}$ \\
    CO & $9.36\times{10}^{-5}$ & $1180$ &  $7\times{10}^{11}$\\
    CO$_2$ & $4.68\times{10}^{-5}$ & $2267$ & $9\times{10}^{11}$\\
    C$_2$H$_6$ & $2.42\times{10}^{-5}$ & $2500$ & $6\times{10}^{16}$\\
    N$_2$ & $3.50\times{10}^{-5}$ & $1051$ & $7\times{10}^{11}$ \\
    NH$_3$ & $7.78\times{10}^{-6}$ & $2715$ & ${10}^{12}$\\
    Ar & $2.78\times{10}^{-6}$ & $866$ &  $6\times{10}^{11}$ \\
    Kr & $1.53\times{10}^{-9}$ & $1371$& ${10}^{14}$ \\
    Xe & $1.91\times{10}^{-10}$ & $1960$ & $4\times{10}^{14}$ \\
    O in refractory solids & $1.87\times{10}^{-4}$ & --&  --\\
    C in refractory solids & $1.45\times{10}^{-4}$ & --& --\\
   S in refractory solids & $1.52\times{10}^{-5}$ & --& --\\
P in refractory solids & $2.95\times{10}^{-7}$ & --&--\\

    \hline
  \end{tabular}
      \begin{tablenotes}
        \raggedright
        \item[a] Abundance calculated from procedures described in \citet{Oberg&Wordsworth19}. We adopt the protosolar elemental abundance recommended by \citet{Asplund+21}.
        \item[b] Desorption energy on compact amorphous water ice: CO and N$_2$from \citet{Fayolle+16}, CO$_2$ from \citet{Noble+12}, C$_2$H$_6$ from \citet{Behmard+19}, NH$_3$ from \citet{Penteado+17} after \citet{Collings+04}, and noble gases from \citet{Smith+16}. For H$_2$O, we adopt the desorption energy from a gold film \citep{Fraser+01}.
    \end{tablenotes}
 \end{threeparttable} 
\end{table}

{ 
\subsection{Atmospheric composition estimation}\label{Appendix:atmosphere}
We estimated the atmospheric elemental abundances of Jupiter and Saturn by summing up disk gas and solid contributions.
Let $\mathcal{N}_{\rm i}$ denote the total number of a species i element in the envelope.
The total number can be written as
\begin{equation}
    \mathcal{N}_{\rm i}=\mathcal{N}_{\rm H}\frac{N_{\rm g,i}}{N_{\rm g,H}}+M_{\rm Z}\frac{q_{\rm d,i}}{m_{\rm i}},
\end{equation}
where $\mathcal{N}_{\rm H}$ is the total number of hydrogen atoms in the envelope and $q_{\rm d,i}=m_{\rm i}n_{\rm d,i}/\sum{m_{\rm i}n_{\rm d,i}}$ is the mass mixing ratio of the species i element in dissolved solids.
We have reasonably assumed that hydrogen comes mostly from disk gases.
Using the envelope mass of $M_{\rm env}\approx m_{\rm H}\mathcal{N}_{\rm H}+m_{\rm He}\mathcal{N}_{\rm He}$, where $\mathcal{N}_{\rm He}$ is the total number of helium atoms in the envelope, we obtain Eq. \eqref{eq:atm_calc} as
\begin{eqnarray}
    \nonumber
    \frac{\mathcal{N}_{\rm i}}{\mathcal{N}_{\rm H}}&=&\frac{N_{\rm g,i}}{N_{\rm g,H}}+\frac{M_{\rm Z}}{M_{\rm env}}\frac{m_{\rm H}\mathcal{N}_{\rm H}+m_{\rm He}\mathcal{N}_{\rm He}}{\mathcal{N}_{\rm H}}\frac{n_{\rm d,i}}{\sum{m_{\rm i}n_{\rm d,i}}},\\
    &=&\frac{N_{\rm g,i}}{N_{\rm g,H}}+\frac{M_{\rm Z}}{M_{\rm env}}\frac{\overline{m}_{\rm H+He}}{\overline{m}_{\rm d}}\frac{\mathcal{N}_{\rm H}+\mathcal{N}_{\rm He}}{\mathcal{N}_{\rm H}}\frac{n_{\rm d,i}}{\sum{n_{\rm d,i}}},
\end{eqnarray}
where $\overline{m}_{\rm H+He}=(m_{\rm H}\mathcal{N}_{\rm H}+m_{\rm He}\mathcal{N}_{\rm He})/(\mathcal{N}_{\rm H}+\mathcal{N}_{\rm He})$ is the mean mass of the hydrogen-helium mixture and $\overline{m}_{\rm d}=\sum{m_{\rm i}n_{\rm d,i}}/\sum{n_{\rm d,i}}$ is the mean mass of the dissolved volatiles.
}
\subsection{Material properties}\label{Appendix:parameter}
Table \ref{table:1} summarizes the total abundances, desorption energy, and desorption attempt frequency of the atoms and molecules considered in this study.
We determined each volatile abundance following the procedure from \citet{Oberg&Wordsworth19} that is motivated by the composition of the interstellar medium.
We set the absolute abundances so that they sum up to the protosolar abundance recommended by \citet[][]{Asplund+21}, as summarized in Table \ref{table:0}.
We partitioned oxygen into refractory silicates (33.33\%), H$_2$O (33.33\%), CO (16.67\%), and CO$_2$ (8.33\%).
The above estimated CO and CO$_2$ abundances comprise about 45\% of the total carbon abundance.
We partitioned the remaining carbons into volatile organics (25\%) and refractory organics (75\%), {where the volatile organics are represented by C$_2$H$_6$, as in \citet{Oberg&Wordsworth19}.}
Nitrogen was partitioned into N$_2$ (90\%) and NH$_3$ (10\%). 
The total abundances of Ar, Kr, and Xe are the same as the protosolar abundances.
We partitioned all S and P into refractory solids.
The actual abundance of each element reservoir is uncertain, as noted by \citet{Oberg&Wordsworth19}. 
However, the detail of each abundance hardly affects the conclusions of this study as all elements are eventually frozen into solids when the disk is so cold that N$_2$ and Ar can freeze.

We adopted the desorption energy for the atoms and molecules adsorbed onto compact amorphous water ices {as H$_2$O is the most abundant ice}.
We used the simultaneously measured attempt frequency, $\nu$, if available; otherwise, we adopted a conventionally used value of $10^{12}~{\rm s^{-1}}$ \citep[e.g.,][]{Penteado+17}.
The desorption energy is potentially higher than that assumed here if the amorphous water ices have porous structures in cold environments \citep[e.g.,][]{Ayotte+01,Fayolle+16}.
In that case, the shadow would further expand the regions where highly volatile substances can freeze.

\section{Mid-plane temperature of dust-rich disks}\label{Appendix:young_disk}
\begin{figure}[t]
\centering
\includegraphics[clip, width=\hsize]{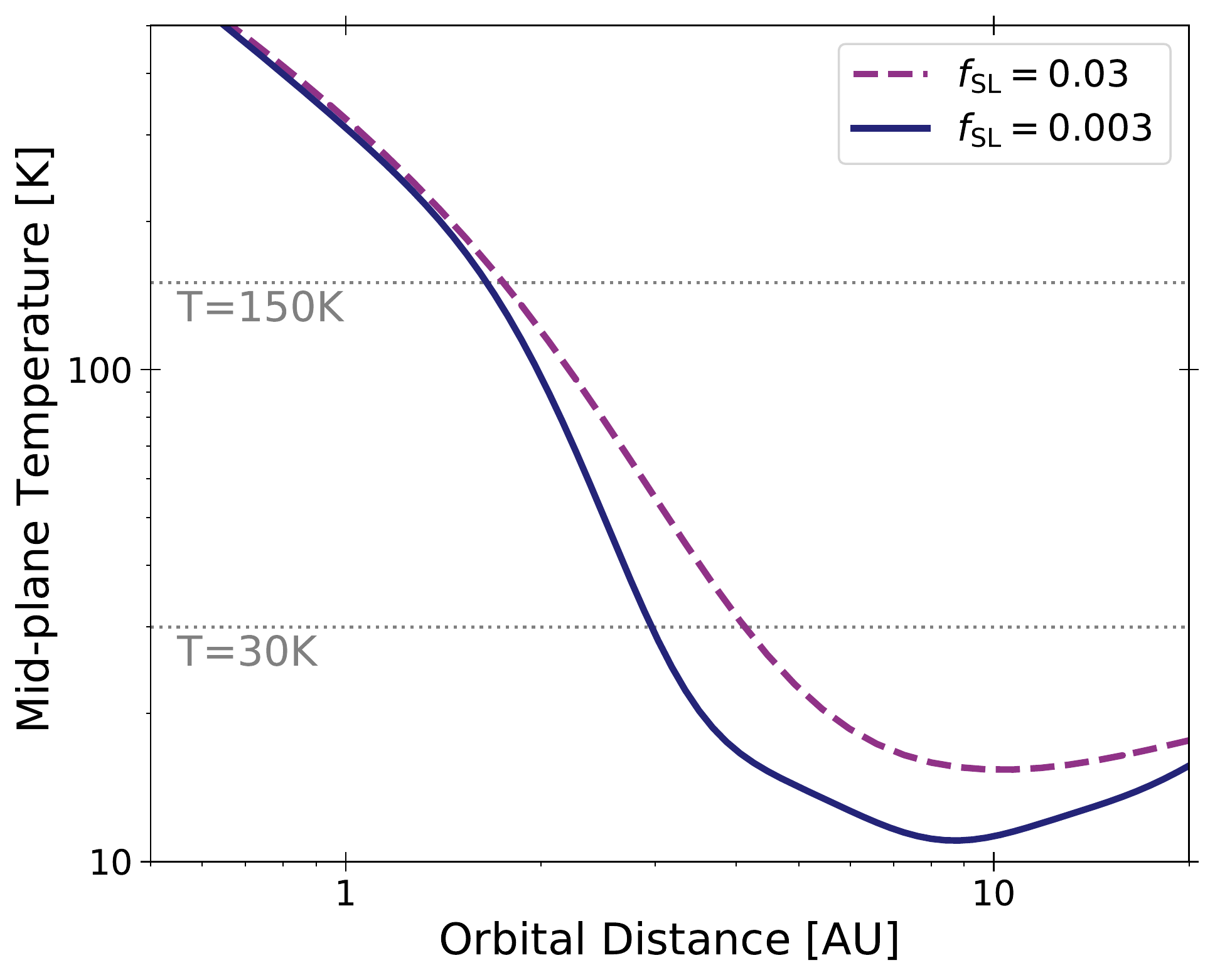}
\caption{Same as Fig. \ref{fig:shadow}, but the dust surface density is an order of magnitude higher than that assumed in Fig. \ref{fig:shadow} (Eq. \ref{eq:appendix_sigmad}).
}
\label{fig:appendix}
\end{figure}
While we have assumed the dust-to-gas mass ratio of $0.01$ inside the H$_2$O snow line, the ratio is likely higher in the younger disks that may be more relevant to the formation of Jupiter. 
Figure \ref{fig:appendix} shows the disk mid-plane temperature for the higher dust surface density of
\begin{equation}\label{eq:appendix_sigmad}
\Sigma_{\rm d} = \left\{ \begin{array}{ll}
    0.1\Sigma_{\rm g} & (r< R_{\rm SL}) \\
    0.1f_{\rm SL}\Sigma_{\rm g} & (r\geq R_{\rm SL}).
  \end{array} \right.
\end{equation}
\ued{Here we conducted the calculations only for $f_{\rm SL}=0.003$ and 0.03. It should be noted that $f_{\rm SL}$ should be less than 0.1; otherwise, the dust-to-gas mass ratio of the whole disk exceeds the interstellar value of 0.01.}
Figure \ref{fig:appendix} \ued{shows} that the shadow cools the vicinity of the current Jupiter orbit to $<30~{\rm K}$ even in the young dust-rich disks.
Thus, we expect that the shadow scenario still holds for relatively young disks.
We note that the snow line moves outward as the dust surface density inside the H$_2$O snow line increases. 
In Fig. \ref{fig:appendix}, we adopt $\alpha_{\rm t}=10^{-4}$ instead of $\alpha_{\rm t}=3\times10^{-4}$ to keep the snow line at $\sim2~{\rm AU}$. 
The actual location of the H$_{2}$O snow line depends on the gas and dust disk properties \citep{Oka+11,Drazkowska&Dullemond18} as well as the accretion mechanism \citep{Mori+19,Mori+21}. 
Therefore, comprehensive parameter studies would be necessary to further understand the shadowing effects on forming giant planets.

\end{document}